\documentclass[3p,times]{elsarticle}

\usepackage{color}

\usepackage{amssymb}

\biboptions{sort&compress}

\usepackage[figuresright]{rotating}

\begin{document}

\begin{frontmatter}

\title{Point contact tunnelling  spectroscopy of the density of states in Tb-Mg-Zn quasicrystals}

\author{R. Escudero}

\ead{escu@unam.mx}

\author{F. Morales\corref{c1}}

\ead{fmleal@unam.mx}

%\cortext[c1]{Corresponding author}

\address{Instituto de Investigaciones en Materiales, Universidad Nacional Aut\'onoma de M\'exico. Apartado Postal 70-360. M\'exico D. F. 04510
M\'EXICO.}

\begin{abstract}
According to theoretical predictions the quasicrystalline (QC) electronic density of states  (DOS) must have a rich and fine spiky structure which actually has resulted elusive. The problem with its absence may be related to poor structural characteristics of the studied specimens, and/or to the non-existence of this spike characteristic. Recent  calculations have shown that the fine structure indeed exists, but only for two dimensional approximants phases. The aim of the present study  is to show our recent experimental studies with point contacts tunnel junction spectroscopy performed in samples of very high quality. The  studies were performed in  icosahedral QC alloys with composition Tb$_9$Mg$_{35}$Zn$_{56}$. We found the presence of a  pseudogap feature at the Fermi level, small as compared to the pseudogap of other icosahedral materials. This study made in different spots on the QC shows quite different spectroscopic features, where the observed DOS was a fine non-spiky structure, distinct to theoretical predictions. In some regions of the specimens the spectroscopic features could be related to Kondo characteristics due to Tb magnetic atoms acting as impurities. Additionally, we observed that the spectroscopic features vanished under magnetic field.

\end{abstract}


\end{frontmatter}

\section{Introduction}

High purity single grain QC specimens obtained in different laboratories \cite{fisher99,fisher98,calvayrac96,boudard95} have shown many of the intrinsic electronic properties related to electronic transport, thermal, and pseudogap characteristics \cite{stadnik96,belin,belin02,rosenbaum00,widmer09,mader13}. Some examples of single grain QCs presenting macroscopically pentagonal faceting, or perfect bulk icosahedral aspect has been obtained by Fisher et al. \cite{fisher99,fisher98}. Electronic transport properties studied in  QCs show  quite different behavior in comparison to the crystalline counterpart. For instance, the electrical resistivity as a function of temperature $\rho(T)$  behaves differently to the crystalline metallic alloys with opposite behavior to the Matthiessen's rule; i.e., the electrical resistivity measured from high to low temperature dramatically increases \cite{belin02,pierce93,kasner95,mayou93,stadnik01}. One of many different physical properties of quasicrystalline alloys and approximants, is that the electronic states have the tendency to be localized at both sides of the Fermi energy, which is a different aspect to the characteristics observed in crystalline intermetallic alloys. The tendency to localization has been related to the  manifestation of the  metallic-insulating transition and the  manifestation of weak electronic localization. In crystalline  intermetallic alloys, the electronic states are extended at both sides of the Fermi energy. Experimental measurements performed on pure and single domain QC structures, show that the ratio between electrical resistances from low to high  temperature  increases with purity and perfection of the QC \cite{fisher99,fisher98,calvayrac96,boudard95,belin02}. Additionally, related to the pseudogap feature, it is important to mention that this characteristic at the Fermi level exist for both crystalline and quasicrystalline alloys, involving a Hume-Rothery stabilization mechanism, it is one of the reasons of the high thermodynamic stability for  icosahedral QCs and  also in  crystalline intermetallic alloys. The only difference is that the pseudogap feature is deeper in QCs than in crystalline alloys \cite{belin02,mizutani98,fujiwara91}.

The existence of the pseudogap has been probed and also the size and shape was determined with great detail using  different spectroscopic techniques; tunnelling, point contacts, and photoemission. \cite{davidov96,escudero99,dolinsek00,banerjee04}. Scanning tunneling spectroscopy measurements taken on Al based QC  have shown spiky local density of states (LDOS) when  the spectroscopy is carried out at the nanometer scale. In  observations at larger surface scale   LDOS obviously is  averaged and  the spiky structure is smeared;  the density of states resembles the total DOS \cite{widmer09,mader13}. It is important to mention that  until today the most notable prediction of the fine spiky structure  has been elusive and no experimental probes have detected it. Accordingly, the general consensus related to  this absence of fine structure is  that the initial theoretical calculations missed relevant information and the resulting spiky  structure was only an artefact of the calculations \cite{fujiwara91}.  The other  relevant aspect could be   related to the  poor QC  characteristics  of  the studied sample \cite{escudero99}. However, until today different arguments  have  been  under different contexts \cite{dolinsek00,zijlstra00,zijlstra03}.

Theoretical studies pointed out that the spiky structure  on DOS  is a characteristic of only small approximants in a two dimensional Penrose tiling and  no related  to the QC  phase. Theoretical studies by Zijlstra and Janssen, and Zijlstra and Bose \cite{zijlstra00,zijlstra03}  resolved structural characteristics at the level of 10 meV  and determined  that spiky structure exists only in approximants phases and not in QC. These  calculations showed   the non-existence spiky structure.

In this investigation  we studied and report point contact tunnelling experiments on Tb$_9$Mg$_{35}$Zn$_{56}$ QC with high quasicrystalline structural order and high purity with second phases estimated below 2-5\% level \cite{fisher98}. Differential conductance as a function of bias voltage shows the existence of a pseudogap. This characteristic depends where the contact spots are placed on the QC surface, in addition, the number of features observed increases as the n-fold surface increases. The results indicate that this pseudogap could be also related to a Kondo behavior or to a spin-glass feature  present in this QC. In the studies carried out  on threefold surfaces, the point contact tunnelling spectra shows fine structure inside the pseudogap at temperatures below the spin-glass freezing temperature.

\begin{figure}
\begin{center}
\includegraphics[scale=0.28]{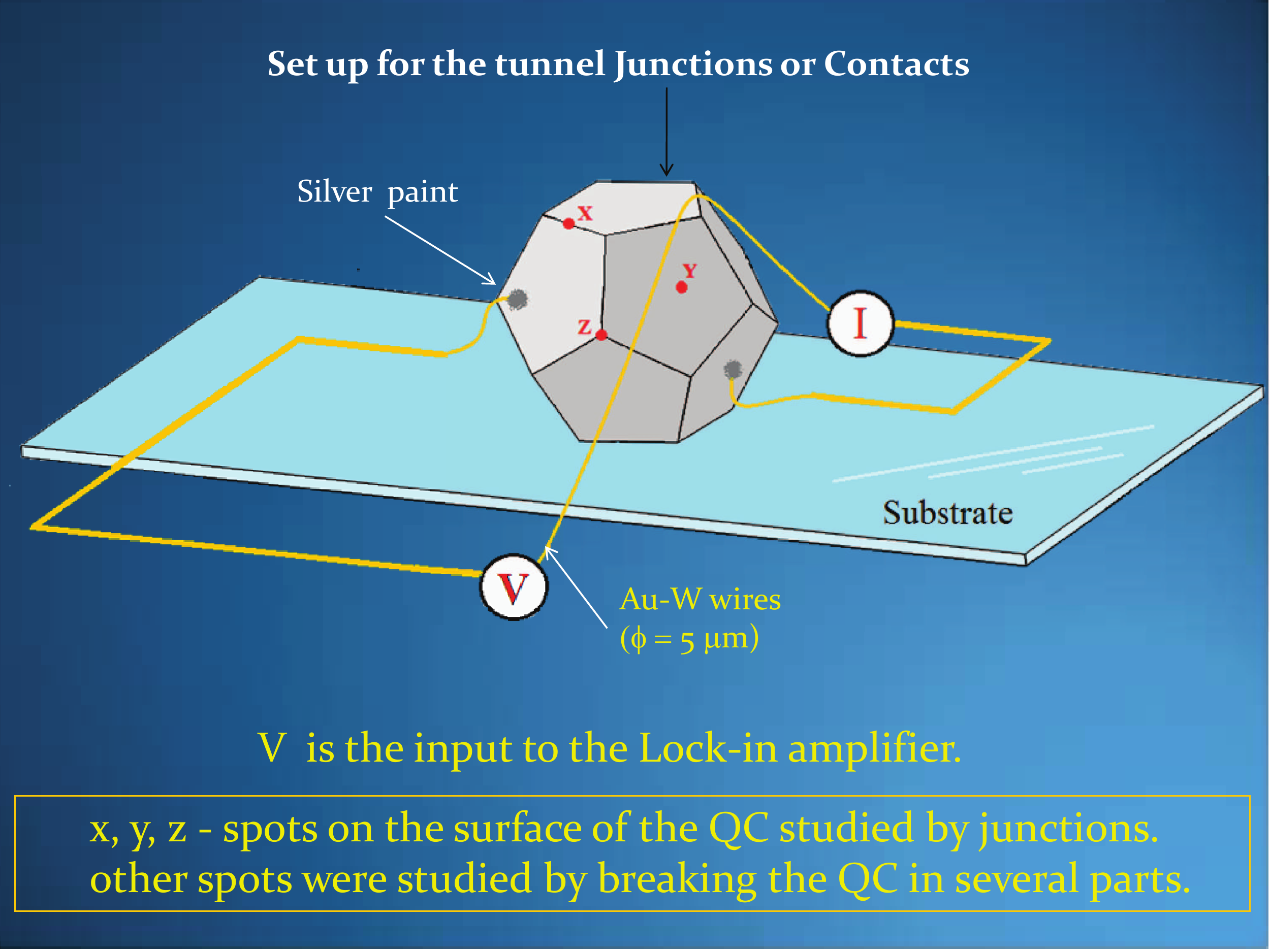}
\caption{\label{fA}(Color online) Set-up for the point contact junctions
in Tb$_9$Mg$_{35}$Zn$_{56}$ an icosahedral quasicrystal studied with Au-W wires to form the junctions on different fold surface spots. Note that
the quasicrystal is represented with the dodecahedral morphology.}
\end{center}
\end{figure}

\section{Experimental details}

The spectroscopic observations in the studied quasicrystals show different structural characteristics on different surfaces  when measured on two, three, and fivefold different regions. The spectroscopic features measured several times on different spots were always similar, considering that the point contact could change  a little of the initial position. Schematically the setup is show  in Fig. \ref{fA}. The determined characteristics were obtained with junctions of areas with few $\mu$m$^2$. The junction size is an important  parameter in tunneling studies. The data collected in this study were the characteristics of very small surfaces. As  the junction area increases fine features could be smoothed and only an average will be observed and the  structure will be  only  the pseudogap feature.

Tunnelling spectroscopy is one of the most sensitive probes to study the density of electronic states. It gives  high spectroscopic resolution limited by temperature, thermal noise is the limiting factor. The thermal energy at 1 K is  86 $\mu$eV. Therefore, with conventional  modulation techniques, the energy resolution will  be of this order of magnitude.

In point contact junctions, the relevant information will be related to scattering and dispersion of the elementary interactions in DOS and also on the features of the relevant spot on the surface. The dV/dI$-$V data were measured at low temperatures with the normal tunnelling electronic, consisting  of lock-in amplifier, bridge, and modulation technique. The modulation consists of a sinusoidal signal at 1000 Hz with amplitude below 50 $\mu$V. Data was stored in a computer and inverted  to obtain the differential conductance dI/dV$-$V. The  spectroscopic data was reproducible and  observed many times in different parts of the QC studied. Tunnelling spectroscopy is a difficult task and much precaution must be taken in order to probe that the  measurement  corresponds  to tunnelling or contacts currents \cite{escudero99}. One possible manner to probe the type of junction is by measuring and  observing the superconducting energy gap when the sample or counter electrode is superconducting. In non-superconducting junctions other tests  could be made, the so-called Rowell criteria. In this criteria, the differential resistance at zero bias as a function of temperature increases as the  temperature decreases \cite{jonsson00}. In a point contact junction the relevant characteristics will be related to pure  Andreev reflections as Blonder et al. model specifies \cite{blonder82}.  In this work the  point contacts  junctions were prepared as explained by Escudero et al. \cite{escudero99}.

The QCs were fabricated and characterised  by Fisher et al. \cite{fisher99,fisher98}. The differential conductance versus energy (voltage) dI/dV$-$V curves show structural features with different widths that change with temperature. The pseudogap feature was about a few mV, smaller than the size reported for Al-Pd-Mn, Al-Cu-Fe, and Al-Pd-Re specimens \cite{davidov96,escudero99}. This small pseudogap could be related  to the small resistances ratio between low and  high temperature as reported by Fisher et al. \cite{fisher99}.

Spectroscopic measurements were performed on different sites of the QC surface, these sites are depicted in Fig.1, furthermore, on regions of undetermined fold sites when the QC was broken. At low temperature  the spectroscopic measurement reveals features, but not of the fine spiky nature.  As the temperature is increased the structure  and pseudogap  becomes   smoothed by  thermal  noise. Determination of the type of junctions, tunnel or metallic  contact, depends of the thickness of the junction insulating barrier. Tunnel junctions  formed with the native oxides grown on the QC  surface sometimes were  good enough to see tunnelling characteristics. In many of the junctions studied the  insulating layer was prepared by cleaning the specimen with diluted acid solution,  washed with distilled water and  finally  exposed  to the air.  The exposed time was variable, from few minutes to hours. Moreover, we have to stress that the exposed time had only limited effect on the growing of the insulating layer, therefore the  junction's differential resistances at zero bias (dV/dI)$_{V=0}$  always presented distinct values. The (dV/dI)$_{V=0}-T$ always increases from high to low temperature, in the range of 10 to 65 Ohm. Junctions with small values  behave as metallic point contact \cite{jonsson00}, with different temperature dependence of the differential resistance at zero bias; those  were analyzed with  the Blonder, Tinkham  and Klapwijk  model \cite{blonder82}. For this study  we prepared more that 200  junction (tunnelling or metallic contacts), many presented reproducible features, those were used for  this work, others were discarded. One important experimental condition, in order to have  minimum averaging on the spectroscopic features, was to keep the junction area as small as possible.

To compare the spectroscopic characteristics of the QCs the dV/dI-V curves of a polycrystalline Tb-Mg-Zn alloy were measured and observed. To obtain the Tb-Mg-Zn polycrystal a small part of the quasicrystal was wrapper with tantalum foil, placed into a quartz tube and encapsulated in argon atmosphere, heated at 700$^\circ$C and maintained at this temperature for 48 hours. After this time the furnace was cooled slowly to room temperature. In this thermal process a small amount of Zn was lost. Additional measurements were performed in a Zn single crystal to compare with the QC.

The  resulting information  in this work was taken with junction formed with the QCs specimen, the Tb-Mg-Zn polycrystal (named  an approximant), and the Zn single crystal, as the first  electrode of  the junction. The second electrode was a thin Au$-$W wire with 5 $\mu$m diameter, and cut in diagonal shape to have small dimension in the tip.  The experimental setup for the junction, allow us to fabricate junctions with areas about  one  $\mu$m$^2$. The setup allows to explore different regions on  the QC surface by  moving the second  electrode to different spots.

\section{Results and discussion}

\begin{figure}[t]
\begin{center}
\includegraphics[scale=0.3]{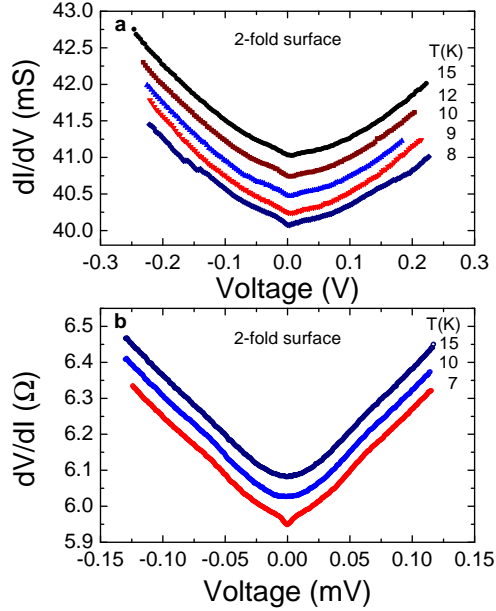}
\caption{(Color online) (a) Differential conductance curves dI/dV versus
voltage of  (Tb$_9$Mg$_{35}$Zn$_{56}$)-(Au-W) tunnel junction measured
from 15 to 8 K on twofold surface spot. At the Fermi level
(zero bias)  the small  depression is the pseudogap and is about 38 mV
size. At 8 K the pseudogap increases to about 55 mV. Note that the
structure is asymmetric around zero bias. (b) Point contact differential
resistance$-$bias voltage measured at 15, 10 and 7 K. The pseudogap on
this twofold surface spot of the QC is only observed at low temperature. In addition, at 7 K the differential resistance shows a tiny structure
about $\pm$50 mV, and the size of the pseudogap feature is only $\sim$16
mV. } \label{f23}
\end{center}
\end{figure}

Spectroscopic data, obtained from a (Tb$_9$Mg$_{35}$Zn$_{56}$)-(Au-W) junction are shown in Fig. \ref{f23}a and \ref{f23}b, these curves were taken on a  twofold surface. The curves shown in Fig. \ref{f23}a are asymmetric, this characteristic is typical of tunneling currents \cite{simmons}. In this figure the pseudogap feature is more pronounced when the temperature  decreases. Note that at 15 K, the feature is small $\pm$ 38 mV and at 8 K the  pseudogap is about $\pm$ 55 mV. Fig. \ref{f23}b shows data of a  contact on a twofold surface measured in  different spot area of Fig. \ref{f23}a. The pseudogap observed at  7 K is $\sim$ 16 mV, small than the observed in Fig. \ref{f23}a. Moreover, two  additional characteristics are  seen, the  differential resistance versus  bias voltage  in the range $\pm  50$ mV  is almost parabolic, and above those  voltages is linear. The curves  at 10 and 15 K do not show the pseudogap feature. The second characteristic  is the symmetry of dV/dI$-$V curves. At this moment we do not have a simple  explanation of this behavior, nevertheless taking into account  that there are not  theoretical models appropriate to  interpret tunnelling or contact measurements on QCs,  the McMillan and Mochel \cite{mcmillan},  and Altshuler et al. \cite{alrshuler79} models give a reasonable fit to the experimental data. These models predict that DOS depresses near the Fermi energy as $\sqrt E$. In Fig. \ref{f23}a, and \ref{f23}b the curves at low temperature between $\pm$50 mV show this energy behavior related to the formation of the pseudogap, and to localization, screening, and correlation effects different to as observed in conducting and crystalline alloys.

\begin{figure}
\begin{center}
\includegraphics[scale=0.3]{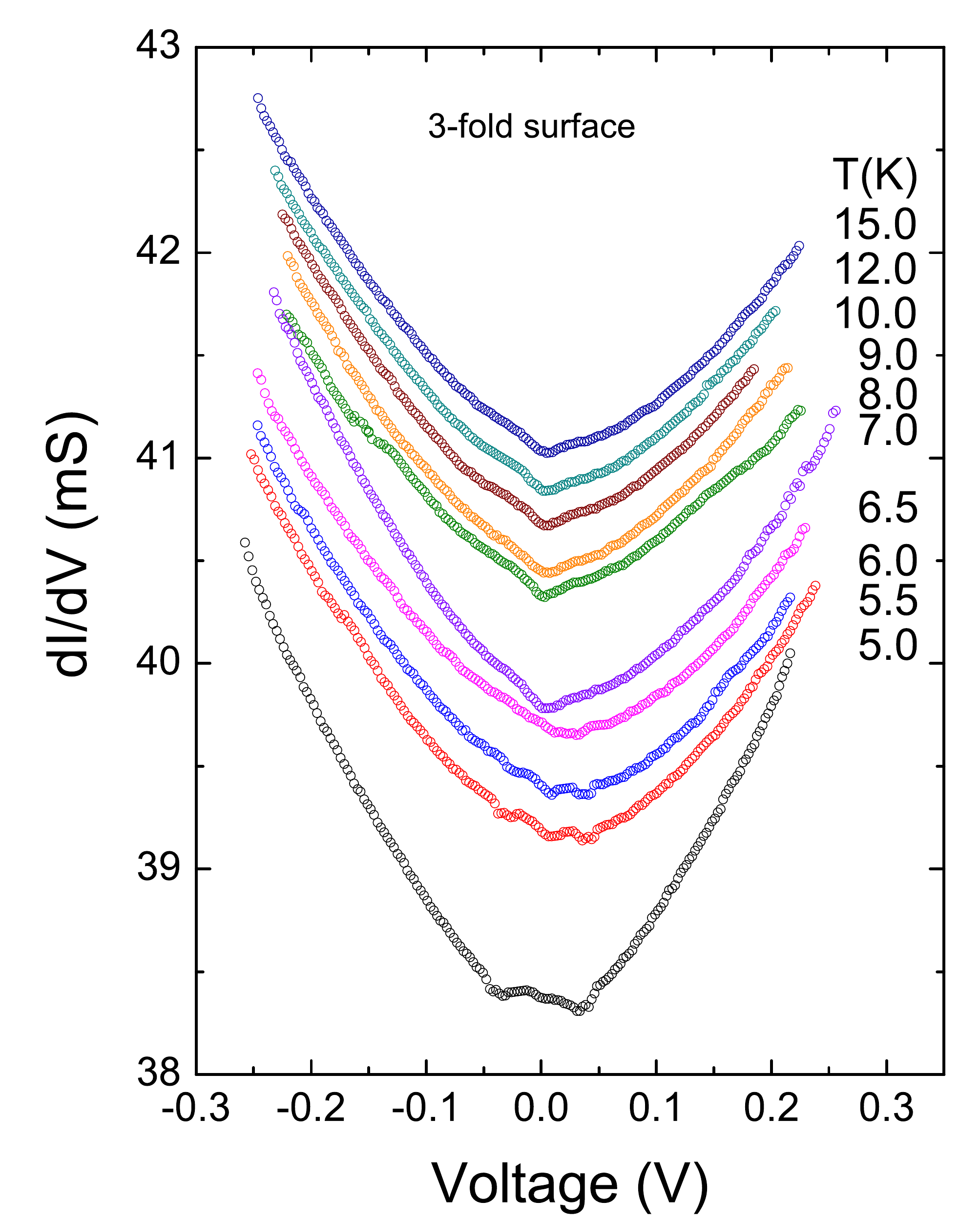}
\caption{\label{f3}(Color online) dI/dV - V determined on a threefold surface spot of the QC. Note  the structure in the pseudogap below 7 K. Above this temperature only a single feature was observed.   At 15 K the pseudogap value is 19 mV. Observe that the total structure from $\pm$200 mV is highly asymmetric around Fermi level.}
\end{center}
\end{figure}

\begin{figure}[h]
\begin{center}
\includegraphics[scale=0.27]{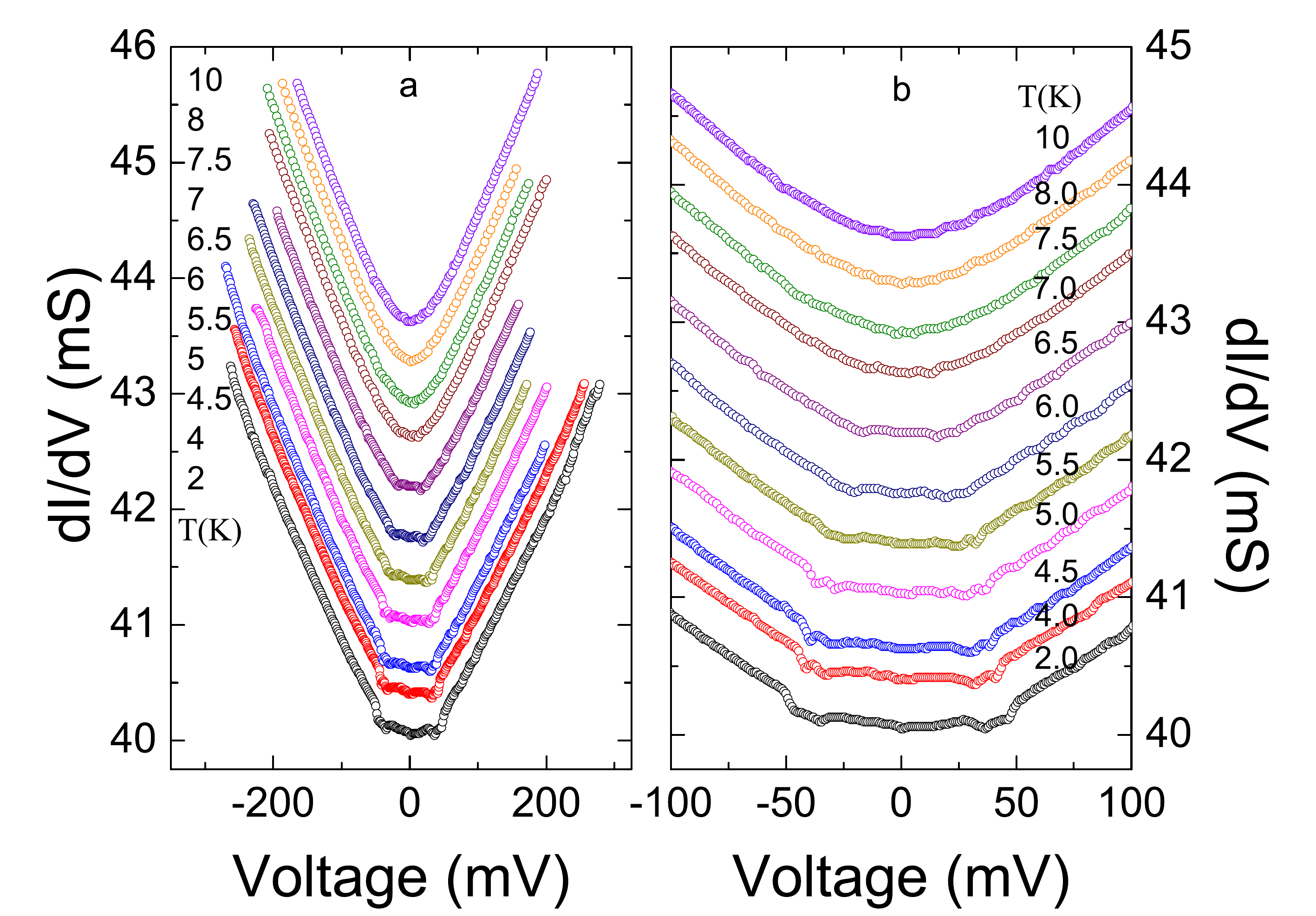}
\caption{\label{f4}(Color online) Differential conductance data {color{red} determined}
on a threefold spot. Panel a shows data measured between
$\pm$300 mV. Note the parabolic shape at high temperature.
Panel b is an amplification of panel a. Structure between $\pm$50 mV, at low temperature is reduced from 2 K up to 7 K.}
\end{center}
\end{figure}

Figures \ref{f3} and \ref{f4} show  differential conductance  dI/dV-V, on two spots of the threefold surface. These figures display curves measured at different temperatures where the temperature evolution of the pseudogap is observed. In Fig. \ref{f3} at 15 K the feature is only a minimum at the zero bias  with  features at high bias, at this temperature the pseudogap size is 19 mV. Decreasing the temperature, dI/dV$-$V  shows fine structure in the interval $\pm$35 mV; the pseudogap  is well defined at low temperature and shows asymmetric structure,  a parabolic shape is well fitted from 15 to 6 K.  Again this is a different behavior as seen in normal metals  and  may be related to the tendency to localization  and increase of screening. Figures \ref{f4}a and \ref{f4}b, show data with similar spectroscopic features in Fig. \ref{f3}. In Fig. \ref{f4}b we present the same data of Fig. \ref{f4}a, but amplified from $\pm$100 mV, note  similar features as in other threefold surface. The  tendency to smoothing the features is clear as the temperature is increased. Fine features can be distinguished from $\pm$50 mV, Fig. \ref{f4}a  shows data at $\pm$300 mV. Junctions characteristics of Fig. \ref{f3} and \ref{f4} were made with Al and Au-W  wires as second electrodes. Both dI/dV$-$V characteristics show similar behavior indicating that this is  independent of the second electrode.

The dI/dV$-$V curves obtained from spots on threefold QC surface show
features that resembled the electronic density of states determined by
theoretical calculations \cite{zijlstra00,zijlstra03} proposed  in a three-dimensional Penrose tiling applied to $\alpha-$AlMn, Al-Pd-Mn QC and approximants.  The main conclusion in these theoretical
studies is related to the absence of spiky fine structure on DOS.

\begin{figure}[h]
\begin{center}
\includegraphics[scale=0.3]{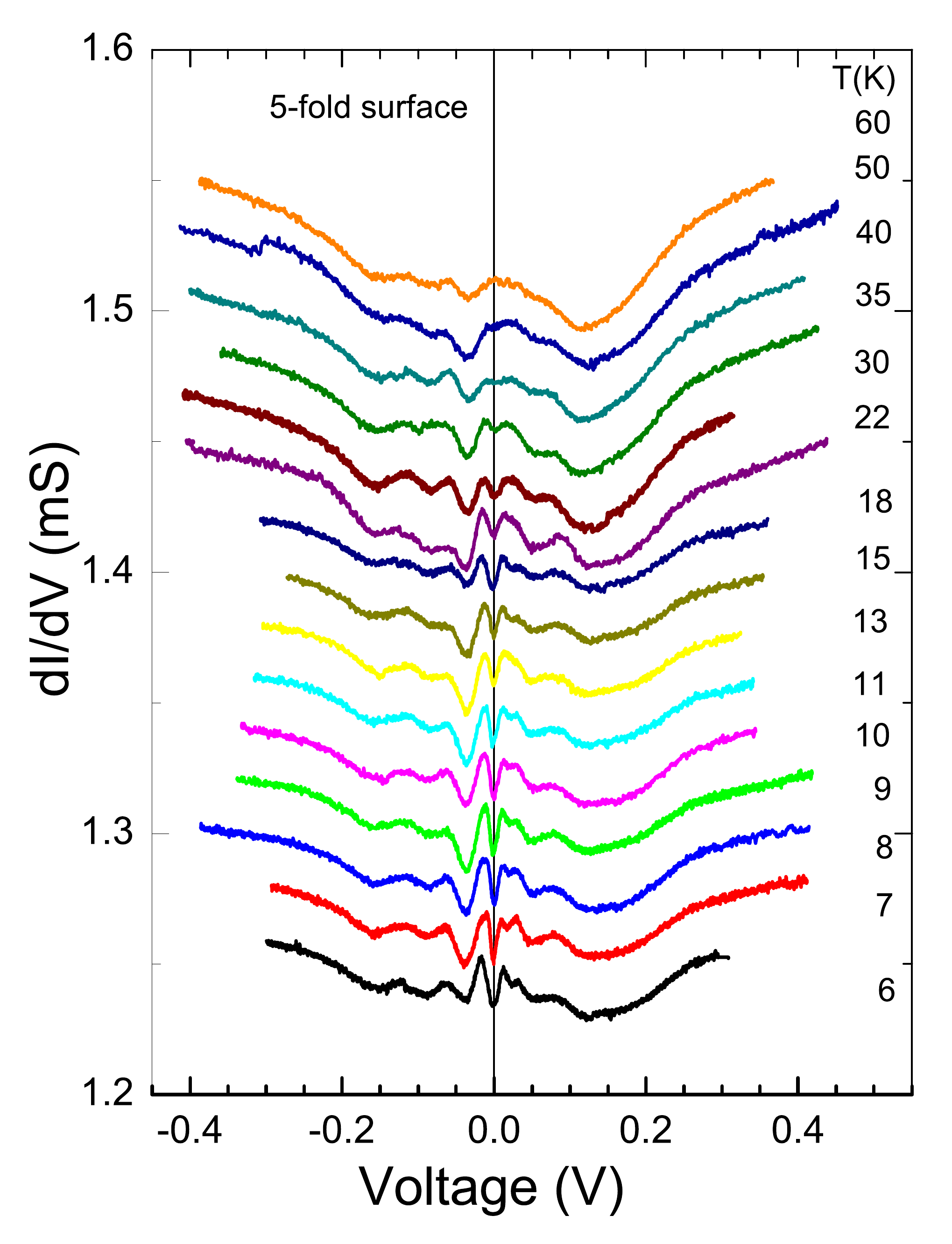}
\caption{\label{f5}(Color online) dI/dV - V data obtained on a fivefold surface. Structure is observed up to $\pm$250 mV.
At Fermi level a pseudogap of $\pm$20 mV is observed, it decreases as the temperature increases. The pseudogap almost disappears above 30 K, and is  absent at 60 K.}
\end{center}
\end{figure}

Data on fivefold spot were  measured and plotted in Fig. \ref{f5} from 6 to 60 K. The characteristics can be seen in the interval  $\pm$300 mV. A small single pseudogap feature is at $\pm$20 mV, decreasing as the temperature is increased. At 35 K the pseudogap is only a very small minimum disappearing at 60 K. At this high temperature, still structure exists in the energy range $\pm$200 mV. The characteristics observed in these curves are quite similar to the dI/dV$-$V curves reported by Widmer et al. \cite{widmer09} and M\"{a}der et al. \cite{mader13}, obtained with scanning tunnelling microscopy in the spectroscopic mode. These measurements were performed on specific places on a small area of a fivefold surface of {\sl i}-Al-Pd-Mn QC. Those authors associate the dI/dV-V curves to  an LDOS. If the curve data is obtained on large surfaces, i. e., $10 \times 10$ nm$^2$, the resulting dI/dV is averaged and the spectra could be considered almost the total DOS \cite{widmer09}. It is noteworthy that similar spiky features were observed on the fivefold surface of different icosahedral QCs. These results  indicate that the fine spiky structure could be related to the fivefold surface.

To confirm the existence of  structure at different regions of the QC, it was  broken to perform  tunnelling  on different parts without knowledge  of the fold region. The features  on  sites of the QC were observed showing distinct  variants depending on the folding area, but similar to the already spots previously measured,  so the  spectral features change depending on   the  different surface fold measured.  It is important to mention that these junctions  gave similar information as already observed; less structural features on two and more on fivefold surface.

In the experimental work reported here, we never observed spiky fine structure as claimed in theoretical studies. We emphasize that the spectroscopic resolution in our junctions is at the order of 86  $\mu$eV, when determined at low temperature with small modulation to detect the differential resistance. More important was the quality of the QCs used \cite{fisher99,fisher98}. We think that this result is  a clear indication that fine spiky structure indeed is nonexisting.  Related to the spectroscopy studied on the  pseudogap, this  was  seen in all different spots, but with small different size.  We believe that this  difference must be related to surface characteristics and different  oxide thickness and imperfections. However, it is important to mention that different junctions evaluated on the same spots  show  similar value of the pseudogap, but different size. So, two conclusions can be obtained; the size of the pseudogap is very sensible to the surface examined and orientation.

\begin{figure}[t]
\begin{center}
\includegraphics[scale=0.7]{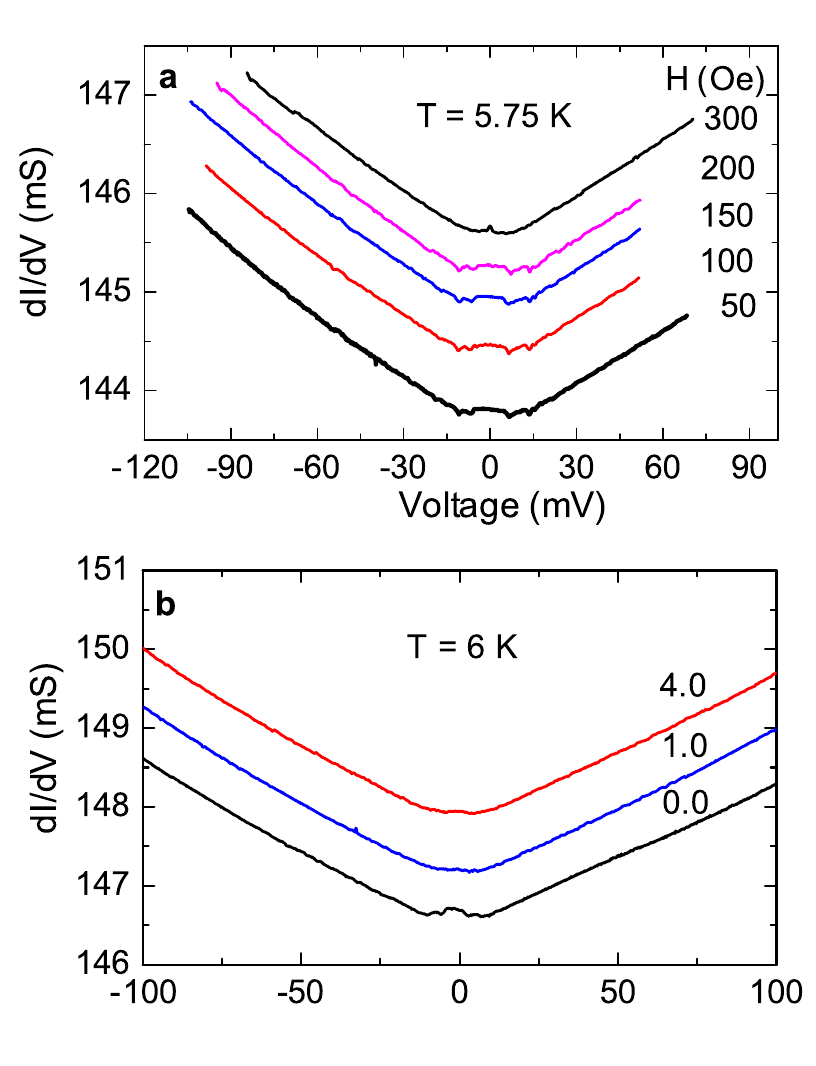}
\caption{\label{f6}(Color online) Effect of the magnetic field on the
differential conductance. Panel a) shows measurements performed at 5.75 K and low applied magnetic field. Panel b) shows the effect of 1 and 4 T over the tunnelling characteristics measured at 6 K. In both cases the spectroscopic features are erased, but at 300 Oe appears only a small zero bias anomaly, meanwhile at higher field a flat region around
zero bias remains.}
\end{center}
\end{figure}

In order to study the influence of  magnetic effects on the QC we
applied magnetic fields in parallel and perpendicular directions to the
plane of the junction,  the morphology of the QC (imperfections, small
size of the sample) was the impediment to precisely determine the QC
directions on the applied magnetic field, nevertheless   both
measurements show similar characteristics and behavior. The structure
around zero bias has a tendency to decrease and to be smoothed as the
magnetic field is increased. Fig. \ref{f6}a and b show details of this
behavior on a threefold surface  with an applied field
perpendicular to the point contact plane. Fig. \ref{f6}a displays  data
at small magnetic fields from 50 to 300 Oe at  $T\sim 5.75$ K, close to the spin glass freezing temperature \cite{Dolinsek}. The pseudogap tends to decrease and shows only  a small zero bias anomaly similar to Appelbaum type  \cite{appelbaum}. It is important to mention that according to Dolin\u{s}et and Jagli\u{c}i\'c \cite{Dolinsek} the spin glass behavior is not conventional, with differences as to a canonical spin glass. However, close to $T\sim 6$ K and a high magnetic field,  the structural features disappear, as  shown  in Fig. \ref{f6}b,  we do not have a plausible physical explanation for this behavior.

\begin{figure}
\begin{center}
\includegraphics[scale=0.28]{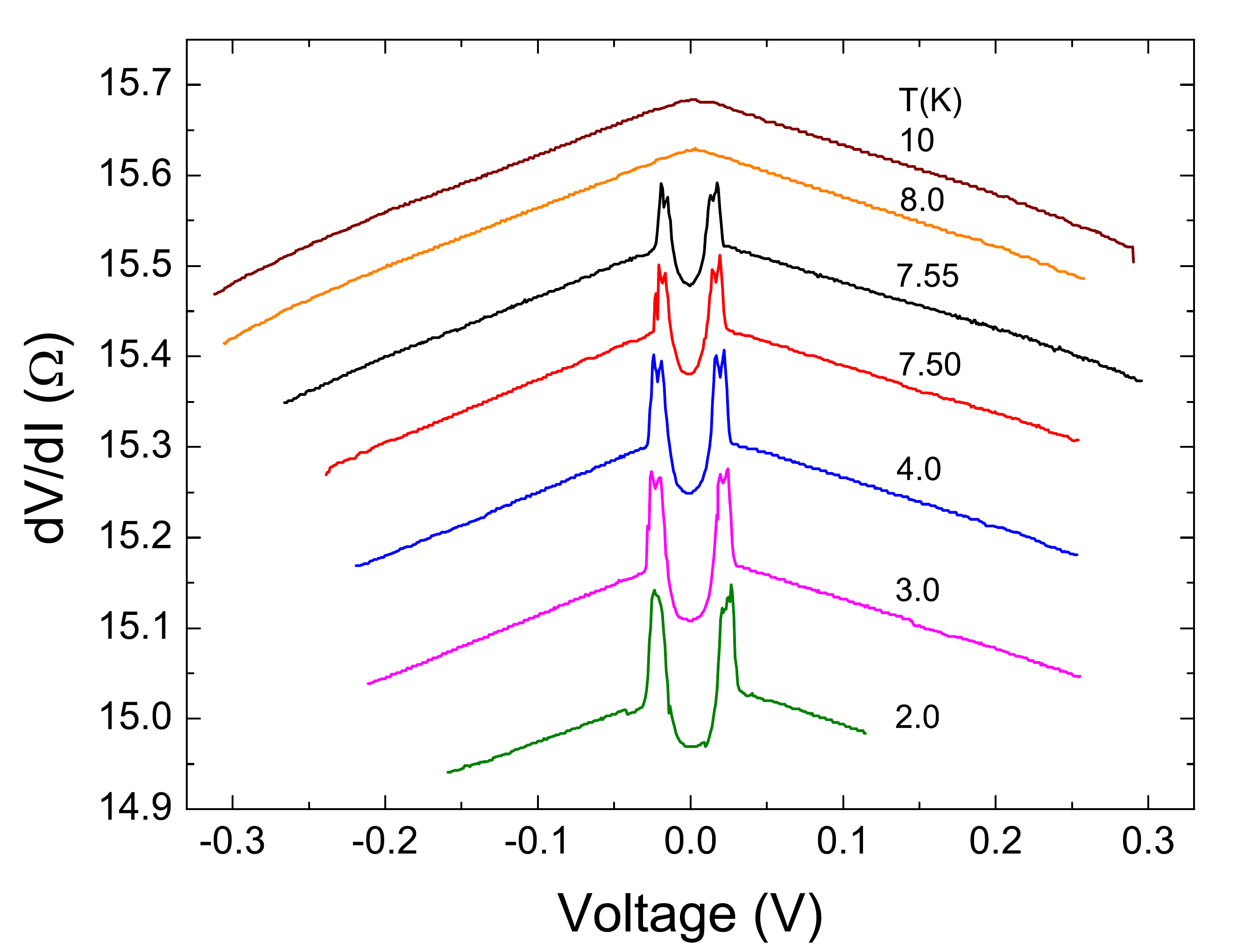}
\caption{\label{f7}(Color online) Point contact differential resistance
data of a spot on the surface of the QC. The feature may be related to a
Kondo characteristic pseudogap, the size of this pseudogap is around
$\pm$  40 mV, probably related to  the spin glass or antiferromagnetic
coupling of Tb-Tb atoms. The temperature at which this feature appears is close to the  spin glass freezing temperature. }
\end{center}
\end{figure}

An interesting piece of information is presented in Fig. \ref{f7}. The curves were obtained on few spots of a new surface when the specimen was broken. The curves show a very well formed pseudogap, with symmetric structure at both sides of the Fermi energy.  A possible explanation of this characteristic may be related to excess of Tb atoms in some non quasicrystalline regions  of the sample. We think that these regions are so infrequent that they do not affect the magnetic behavior of the QC. The features are  quite similar to  the signature of Kondo effect, occurring by a possible hybridization gap \cite{park} which arises by the influence of Tb $f$ localized electrons and $d$ conduction electrons. The  size of this energy gap is $\pm 40$ mV, it decreases as  temperature increases disappearing at 8 K.

Lastly, we mention that the magnetic characteristics of
Tb$_9$Mg$_{35}$Zn$_{56}$ QC  exhibits  a Curie-Weiss type  behavior with
negative Weiss temperature, indicative of antiferromagnetic exchange
interaction between the Tb atoms. Accordingly, this behavior  has been
related to a spin glass characteristics presented below 5.8 K,  the
spin-glass freezing temperature \cite{Dolinsek,fisher99}.

\begin{figure}
\begin{center}
\includegraphics[scale=0.3]{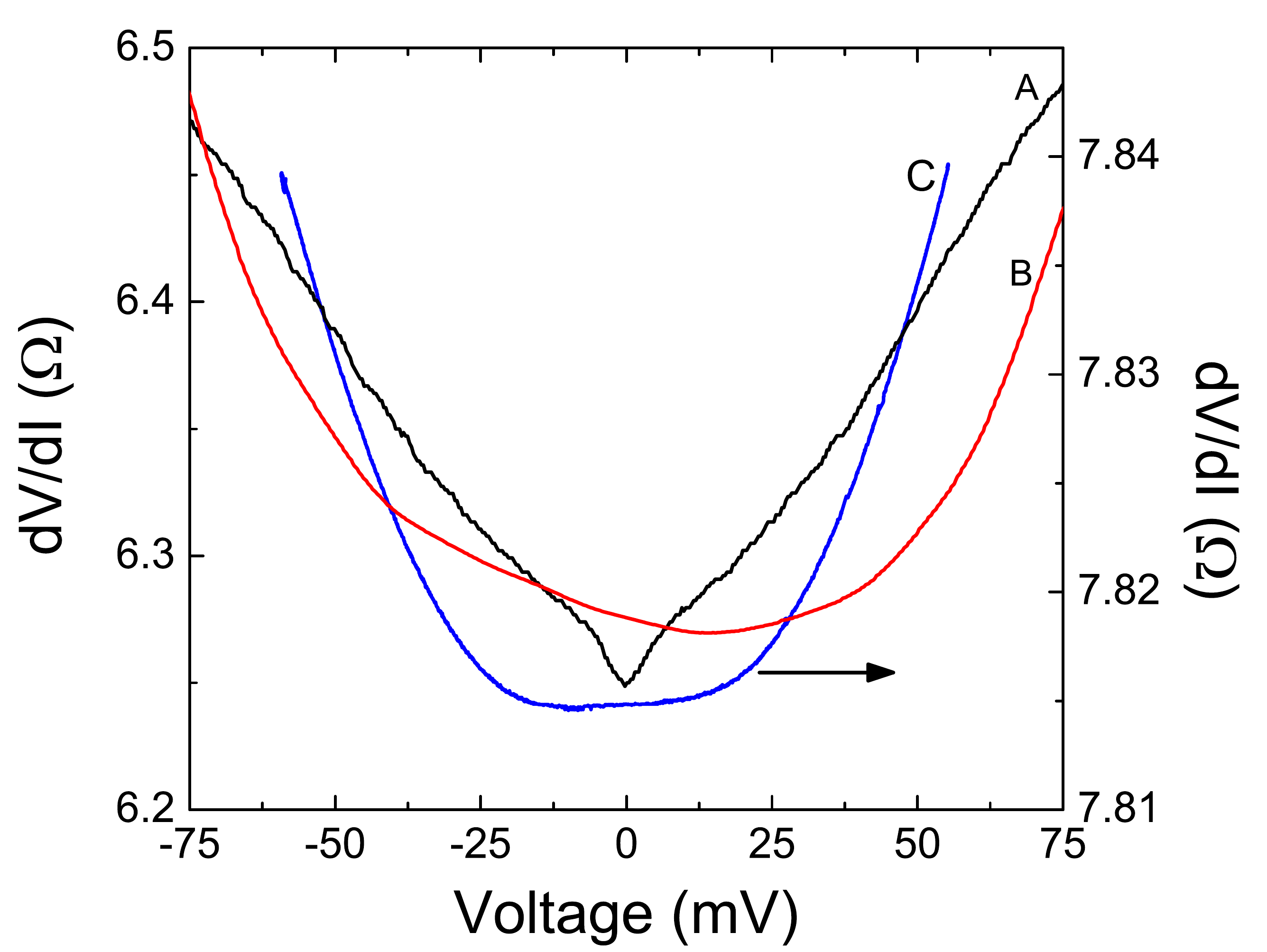}
\caption{\label{f9}(Color online) Point contact differential resistance
as a function of bias voltage of three different samples. Curve A is from a Tb$_9$Mg$_{35}$Zn$_{56}$-(Au-W) point contact measured at 7 K. Curve B
is from a Tb-Mg-Zn polycrystal and gold wire, measured at 10 K. Curve C
is from a Zn single crystal and gold wire, measured at 5 K. Note the
depression around zero bias in the quasicrystal that we assigned as a
pseudogap, it is not observed in curves B and C. The arrow indicates the
vertical scale associated to curve C. Curve B was shifted up for an easy
comparison with other curves.}
\end{center}
\end{figure}

Figure \ref{f9} shows the curves of the already measured Tb$_9$Mg$_{35}$Zn$_{56}$ QC, curve A, Tb-Mg-Zn polycrystal sample curve B, and Zn single crystal curve C. In this figure curve A is the same as of Fig. 2b, QC measured at 7 K. Curves B and C, were measured at 10 K and 5 K respectively. It is noteworthy that in curve A clearly may be seen at zero bias the diminishing of the differential resistance, interpreted as the pseudogap feature of the QC. Measurements of Tb-Mg-Zn polycrystal and Zn single crystal do not show the pseudogap feature, only show a flat region around zero bias. Those last experiments, shown in Fig. \ref{f9}, were performed in order to demonstrate that the pseudogap observed is an electronic characteristic of Tb$_9$Mg$_{35}$Zn$_{56}$ QC, this fact is in agreement with similar experiments reported for other QCs \cite{davidov96,escudero99,dolinsek00,banerjee04}.

\section{Conclusions}

We have demonstrated that a fine  structure on the electronic density states indeed exists in Tb$_9$Mg$_{35}$Zn$_{56}$ in  spots of the QC studied. This experimental work shows different behavior  to the theoretical predictions of the existence of a spiky fine structure. The spectroscopic features determined at the different fold surfaces show that the number of structural features increase as the fold symmetry increases. Twofold spots present only a small pseudogap, moreover  as the  n-fold increases more  features appear. The  Structure inside of the pseudogap is according  to the theoretical studies by  Zijlstra, et al. \cite{zijlstra00,zijlstra03}. In those theoretical  studies they determined a pseudogap feature at the Fermi level using  an ideal three dimensional Penrose tiling  for the  approximant and for a QC with compositions $\alpha -$AlMn and Al-Pd-Mn. Lastly,  as a conclusion of our study we addressed  that the study  on QC needs more experimental and  theoretical work in order to have better understanding, and perhaps these   experimental results in Tb-Mg-Zn may be useful to improve this  interesting field.

\section*{Acknowledgements}
The single grain specimens were kindly provided by Prof. I. R. Fisher of
Stanford University. We thank interesting comments by  E. Maci\'a,  E. S. Zijlstra, E. Belin-Ferre, and D. Mayou. RE thanks to CONACyT and
DGAPA-UNAM project  IN106014.

\end{document}